\documentclass[a4paper,fleqn,usenatbib]{mnras}

\usepackage{newtxtext,newtxmath}
\usepackage{mathptmx}
\usepackage{txfonts}

\usepackage[T1]{fontenc}
\usepackage{ae,aecompl}
\usepackage{textcomp, libertine}
\usepackage{graphicx}	
\usepackage{amsmath}	
\usepackage{amssymb}	
\usepackage[T1]{fontenc} 
\usepackage{aecompl}

\title[The complex field of $\epsilon$ Eridani at activity minimum]{The relation between stellar magnetic field geometry and chromospheric activity cycles I: The highly variable field of $\epsilon$ Eridani at activity minimum}

\author[S.V.Jeffers et al.]{
S.~V.~Jeffers,$^{1}$
S.~Boro Saikia$^{1}$
J.~R.~Barnes,$^{2}$
P.~Petit,$^{3,4}$
S.~C.~Marsden,$^{5}$ \newauthor
M.~M.~Jardine,$^{6}$
A.~A.~Vidotto,$^{7}$
and the BCool collaboration \thanks{Based on observations made through OPTICON with Telescope Bernard Lyot (TBL, Pic du Midi, France) of the Observatoire Midi-Pyrenees, which is operated by the Institut National des Sciences de l'Univers of the Centre National de la Recherche Scientifique (CNRS) of France.}
\\
$^{1}$Institut f\"{u}r Astrophysik, Georg-August-Universit\"{a}t, Friedrich-Hund-Platz 1, 37077 G\"{o}ttingen, Germany \\
$^{2}$School of Physical Sciences, The Open University, Walton Hall, Milton Keynes MK7 6AA, UK  \\
$^{3}$Universit\'{e} de Toulouse, UPS-OMP, Institut de Recherche en Astrophysique et Plan\'{e}tologie, Toulouse, France  \\
$^{4}$Institut de Recherche en Astrophysique et Plan\'{e}tologie, 14 Avenue Edouard Belin, F-31400 Toulouse, France  \\
$^{5}$Computational Engineering and Science Research Centre, University of Southern Queensland, Toowoomba, 4350, Australia  \\ 
$^{6}$SUPA, School of Physics and Astronomy, University of St Andrews, North Haugh, St Andrews, Fife KY 16 9SS, UK  \\
$^{7}$School of Physics, Trinity College Dublin, the University of Dublin, Dublin-2, Ireland
}

\date{Accepted XXX. Received YYY; in original form ZZZ}

\pubyear{2016}

\begin{document}
\label{firstpage}
\pagerange{\pageref{firstpage}--\pageref{lastpage}}
\maketitle

\begin{abstract}
The young and magnetically active K dwarf $\epsilon$ Eridani exhibits a chromospheric activity cycle of about 3 years.  Previous reconstructions of its large-scale magnetic field show strong variations at yearly epochs.  To understand how $\epsilon$ Eridani's large-scale magnetic field geometry evolves over its activity cycle we focus on high cadence observations spanning 5 months at its activity minimum.  Over this timespan we reconstruct 3 maps of $\epsilon$ Eridani's large-scale magnetic field using the tomographic technique of Zeeman Doppler Imaging.  The results show that at the minimum of its cycle, $\epsilon$ Eridani's large-scale field is more complex than the simple dipolar structure of the Sun and 61 Cyg A at minimum.  Additionally we observe a surprisingly rapid regeneration of a strong axisymmetric toroidal field as $\epsilon$ Eridani emerges from its S-index activity minimum.  Our results show that all stars do not exhibit the same field geometry as the Sun and this will be an important constraint for the dynamo models of active solar-type stars.
\end{abstract}




\section{Introduction} 

The evolution of the Sun's large-scale magnetic field ranges from dipolar at activity minimum to complex at activity maximum ~\cite{Derosa2012}.  A solar-like magnetic cycle has also been observed in the K-dwarf 61 Cyg A, where its large-scale field is a simple dipole at activity minimum.  In this paper we investigate the evolution of the large-scale magnetic field of $\epsilon$ Eridani which is well established to be a magnetically active star ~\cite{Valenti1995} and \cite{Metcalfe2013}.  Previously in ~\cite{Jeffers2014}, we reconstructed the large-scale magnetic field geometry of $\epsilon$ Eridani to understand how the photospheric large-scale magnetic field geometry of $\epsilon$ Eridani varies over its S-index cycle.   These observations comprise 6 epochs spanning nearly 7 years or approximately 2 S-index cycles.  We showed that each map has evolved dramatically from one epoch to the next, and that we clearly reconstruct the weakest magnetic field structures at its Ca II H\&K (or S-index) minimum.  The motivation for this present work is to investigate the evolution of $\epsilon$ Eridani's large-scale magnetic field with a higher cadence of observations over its S-index minimum to understand how its field evolution differs from the Sun and 61 Cyg A.  To achieve this, we obtained spectropolaritmetric observations every night, weather permitting, over a period of five months.   

\begin{table}
\caption{Stellar Parameters}
\protect\label{t-stparam}
\begin{tabular}{l c c }
\hline
\hline
Parameter & Value & Reference\\
\hline
Magnitude &  V=3.7 \\
Spectral Type & K2V & \cite{Valenti2005}\\
Distance & 3.2 pc & \cite{vanLeeuwen2007} \\
Effective Temperature  & 5146$\pm$31 K & \cite{Valenti2005} \\
Mass (M$_\odot$) & 0.856$^{+0.006}_{-0.008}$  & \cite{Valenti2005} \\ 
Radius (R$_\odot$) & 0.74$\pm$0.01 &  \cite{Baines2012} \\
$v$ sin $i$ (km s$^{-1}$) & 2.2 $^{+0.04}_{-0.04}$ & \cite{Brewer2016} \\
P$_\mathrm{rot}$ (day) & 11.68 & \cite{Donahue1996} \\
Inclination & 46$\pm 2^\circ$ & \cite{Jeffers2014} \\
Age  & 440 Myr & \cite{Barnes2007}\\
\hline
\hline
\end{tabular}
\end{table}

\section{Observations and Data Analysis}

We observed $\epsilon$ Eridani over a time span of 5 months from September 2014 to January 2015 using the high-resolution spectropolarimeter NARVAL located at the Telescope Bernard Lyot, France \citep{Auriere2003}.   The total data set comprises 40 spectra which were obtained every night with acceptable observational conditions and are summarised in Table~\ref{t-obslist}. The data were reduced and processed following an identical procedure already explained in Section 3 of \cite{Jeffers2014}. 

\begin{table}
\caption{Journal of Observations.  Phase = 0 is defined as Julian date = 2454101.5 and is used for all epochs, with subsequent epochs taking phase = 0 as an integer number of rotational periods from this value. The exposure time of all observations is 400s.}
\protect\label{t-obslist}
\begin{tabular}{l c c c c c }
\hline
\hline
Date & Julian Date & UT & Phase & LSD & S-index\\
& (+2454000) &&& S/N &\\
\hline
\multicolumn{6}{c}{Map 1 (2014.71)} \\
01sep14 & 2902.67 & 04:05:45 & -1.1737 &  39478 & 0.402 \\
02sep14 & 2903.68 & 04:24:48 & -1.0870 &  36435 & 0.396 \\
03sep14 & 2904.67 & 04:00:01 & -1.0029 &  37224 & 0.395 \\
05sep14 & 2906.65 & 03:38:55 & -0.8329 &  42847 & 0.394 \\
11sep14 & 2912.65 & 03:33:12 & -0.3195 &  35636 & 0.404 \\
12sep14 & 2913.62 & 02:53:49 & -0.2362 &  38801 & 0.399 \\
13sep14 & 2914.66 & 03:47:37 & -0.1474 &  39293 & 0.396 \\
23sep14 & 2924.64 & 03:16:40 & 0.7069 &  38855 & 0.405 \\
24sep14 & 2925.6 & 02:19:16 & 0.7891 &  36714 & 0.396 \\
25sep14 & 2926.62 & 02:52:49 & 0.8767 &  35734 & 0.403 \\
27sep14 & 2928.61 & 02:36:02 & 1.0469 &  35976 & 0.391 \\
\\
\multicolumn{6}{c}{Map 2 (2014.84)} \\
16oct14 & 2947.65 & 03:36:25 & -1.3227 &  30832 & 0.396 \\
17oct14 & 2948.49 & 23:49:35 & -1.2506 &  34626 & 0.395 \\
18oct14 & 2949.63 & 03:09:04 & -1.1531 &  36745 & 0.392 \\
19oct14 & 2950.58 & 01:59:09 & -1.0717 &  28518 & 0.393 \\
24oct14 & 2955.48 & 23:37:39 & -0.6520 &  29839 & 0.405 \\
25oct14 & 2956.56 & 01:23:13 & -0.5601 &  36520 & 0.403 \\
26oct14 & 2957.44 & 22:40:33 & -0.4842 &  24285 & 0.410 \\
28oct14 & 2959.59 & 02:07:25 & -0.3006 &  33182 & 0.402 \\
29oct14 & 2960.53 & 00:47:06 & -0.2198 &  36670 & 0.401 \\
30oct14 & 2961.49 & 23:39:16 & -0.1382 &  33682 & 0.397 \\
31oct14 & 2962.51 & 00:13:05 & -0.0506 &  36824 & 0.401 \\
01nov14 & 2963.51 & 00:10:47 & 0.0349 &  42832 & 0.403\\
12nov14 & 2974.53 & 00:45:23 & -1.0213 &  20765 & 0.388 \\
15nov14 & 2977.48 & 23:38:01 & -0.7684 &  31998 & 0.309 \\
20nov14 & 2982.45 & 22:45:26 & -0.3435 &  21015 & 0.387 \\
\\
\multicolumn{6}{c}{Map 3 (2014.98)} \\
02dec14 & 2994.39 & 21:21:14 & -2.3211 &  28400 & 0.387 \\
02dec14 & 2994.40 & 21:32:10 & -2.3204 &  29646 & 0.387 \\
03dec14 & 2995.40 & 21:30:49 & -2.2349 &  38856 & 0.390 \\
03dec14 & 2995.40 & 21:41:47 & -2.2342 &  38417 & 0.390 \\
19dec14 & 3011.39 & 21:17:49 & -0.8658 &  42049 & 0.385 \\
19dec14 & 3011.39 & 21:28:46 & -0.8652 &  39688 & 0.384 \\
21dec14 & 3013.40 & 21:38:44 & -0.6933 &  39051 & 0.392 \\
21dec14 & 3013.41 & 21:49:41 & -0.6927 &  37356 & 0.392 \\
06jan15 & 3029.36 & 20:35:57 & 0.6728 &  39935 & 0.390\\
06jan15 & 3029.37 & 20:46:56 & 0.6735 &  39952 & 0.389 \\
10jan15 & 3033.36 & 20:36:39 & 1.0153 &  41780 & 0.384 \\
10jan15 & 3033.37 & 20:47:36 & 1.0160 &  41215 & 0.385 \\
17jan15 & 3040.33 & 19:56:13 & 1.6122 &  31182 & 0.380 \\
17jan15 & 3040.34 & 20:07:11 & 1.6129 &  32611 & 0.379 \\
\hline
\hline
\end{tabular}
\end{table}


All {\em{Stokes I}} and {\em{Stokes V}} reduced spectra were processed using LSD \citep[Least-Squares Deconvolution][]{Donati1997}.  By extracting the information contained in each spectral line, LSD enables the recovery of weak signatures in the line profile despite the presence of noise.   The result is one high signal-to-noise spectral profile for each observation.  To achieve this, a line list for a star with: $T_{\rm eff}=5000$K, $\log g=4.5$, a depth threshold of 0.1 and $\log({\rm M/H})=0.1$ and containing 12220 lines was downloaded from the VALD database.  This model atmosphere was used to generate a spectral weighting mask.  This mask was applied to each observation (comprising a subset of 4 exposures with different polarimeter angles) with a step size of 1.8 km s$^{-1}$, matching NARVAL's detector pixel resolution.  The resulting Stokes V profiles, which when convolved with the model line profile best matches the observed spectrum of the target star is shown in the outer panels of Figure~\ref{f-magmaps}.  To parameterise the amount of emission in the Ca II H\&K values we use an S-index calculated as described in \cite{Jeffers2014}, where the S-index values calculated using NARVAL spectra are calibrated to the values from the Mount Wilson S-index survey by \cite{Marsden2014}.  

\section{Large-scale magnetic field geometry}

The large-scale magnetic field geometry is reconstructed using the tomographic technique of Zeeman-Doppler imaging which incorporates the maximum entropy algorithm described by \cite{skilling1984}.  This method uses the stellar parameters shown in Table 1 to model local {\em{Stokes V}} profiles sampled over the stellar surface from which a disk integrated synthetic {\em{Stokes V}} profile is computed.  This is then used to iteratively fit the model {\em{Stokes V}} profiles to the observed {\em{Stokes V}} profiles.  The tomographic images of the large-scale magnetic field topology of $\epsilon$ Eridani are reconstructed by assuming that the field geometry is projected onto a spherical harmonics frame \citep{Donati2006}, where the magnetic energy is decomposed into poloidal and toroidal components.  A spherical harmonics expansion with $\ell_{max}=10$ was used as there was no improvement to the fits using larger values.  A reduced $\chi ^2$ of 1.05 was obtained for all of the maps when differential rotation was included in the image reconstruction process.

\subsection{Magnetic maps}

\begin{figure*}
\def\imagetop#1{\vtop{\null\hbox{#1}}}
\begin{tabular}[h]{p{3cm} p{5cm} p{5cm} p{3cm}} 
  \emph{\hspace{0.4cm} Stokes V profiles} & \emph{\hspace{2.25cm} Magnetic maps} & \emph{\hspace{1.55cm} Magnetic maps} & \emph{\hspace{0.4cm} Stokes V profiles}\\ 
  \imagetop{\includegraphics[height=7.5cm,width=3.4cm,bb = 54 -36 520 780,clip]{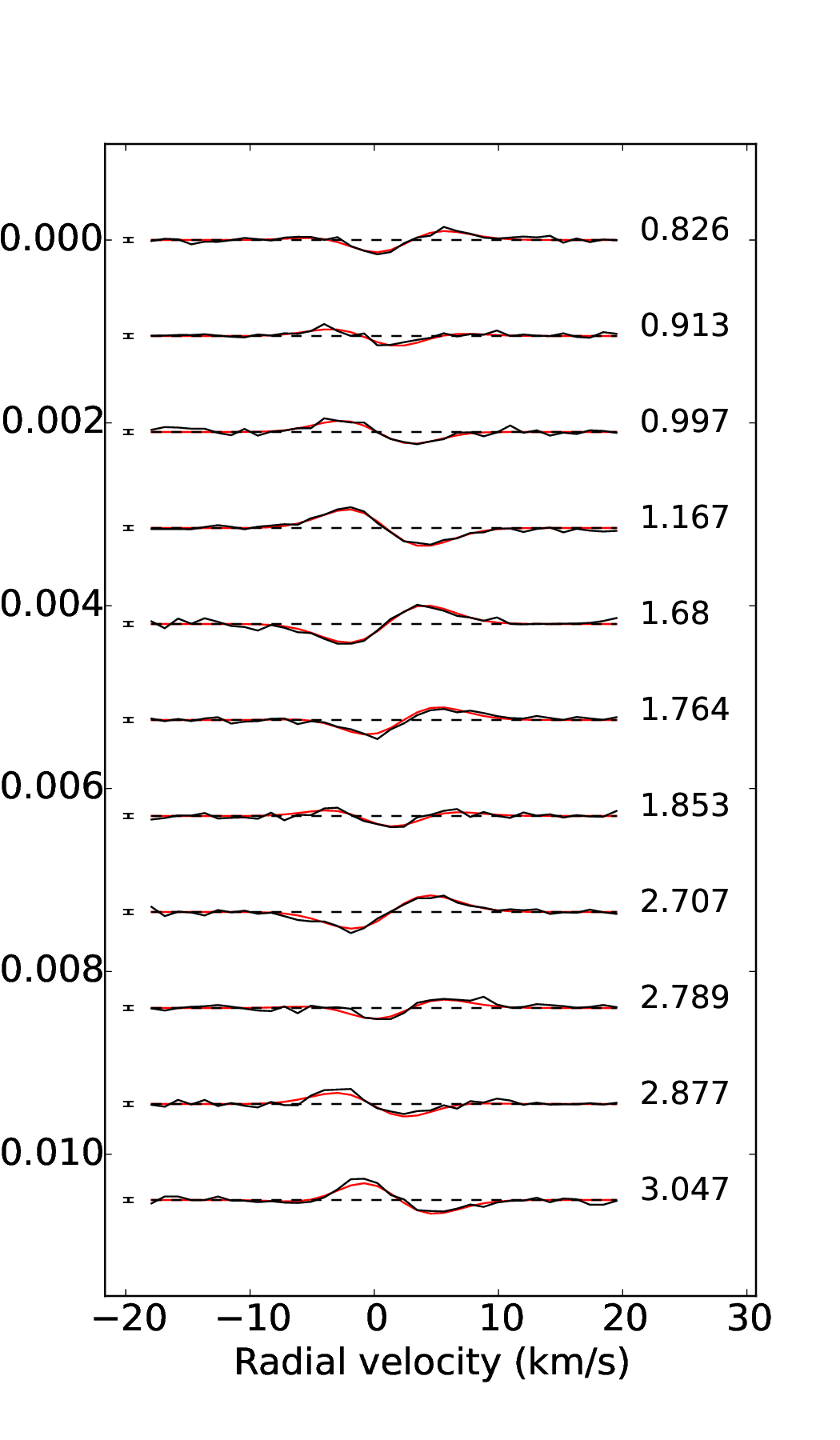}} &   
  \imagetop{\includegraphics[height=7.0cm,width=0.29\textwidth]{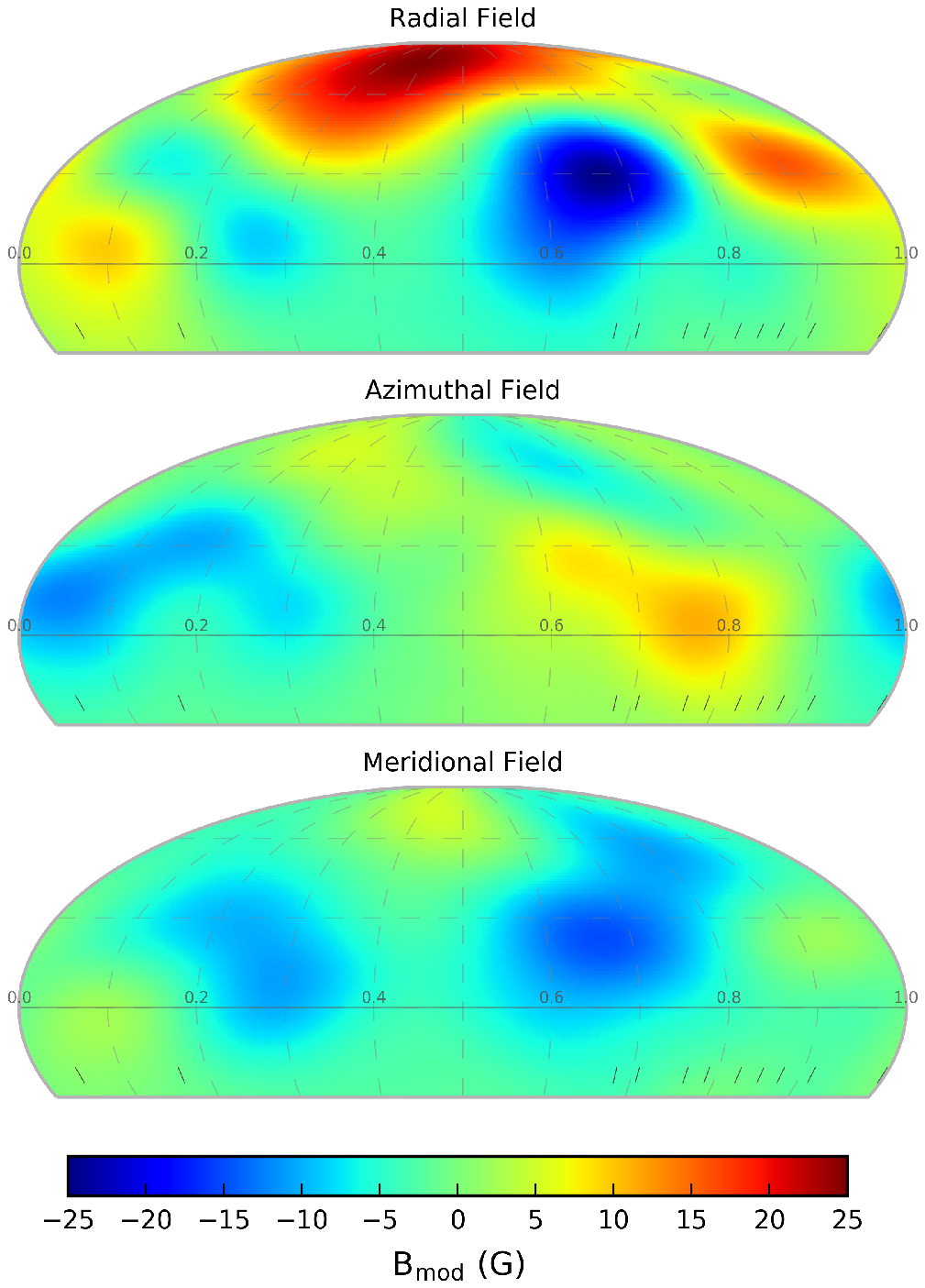}} &
  \imagetop{\includegraphics[height=7.0cm,width=0.29\textwidth]{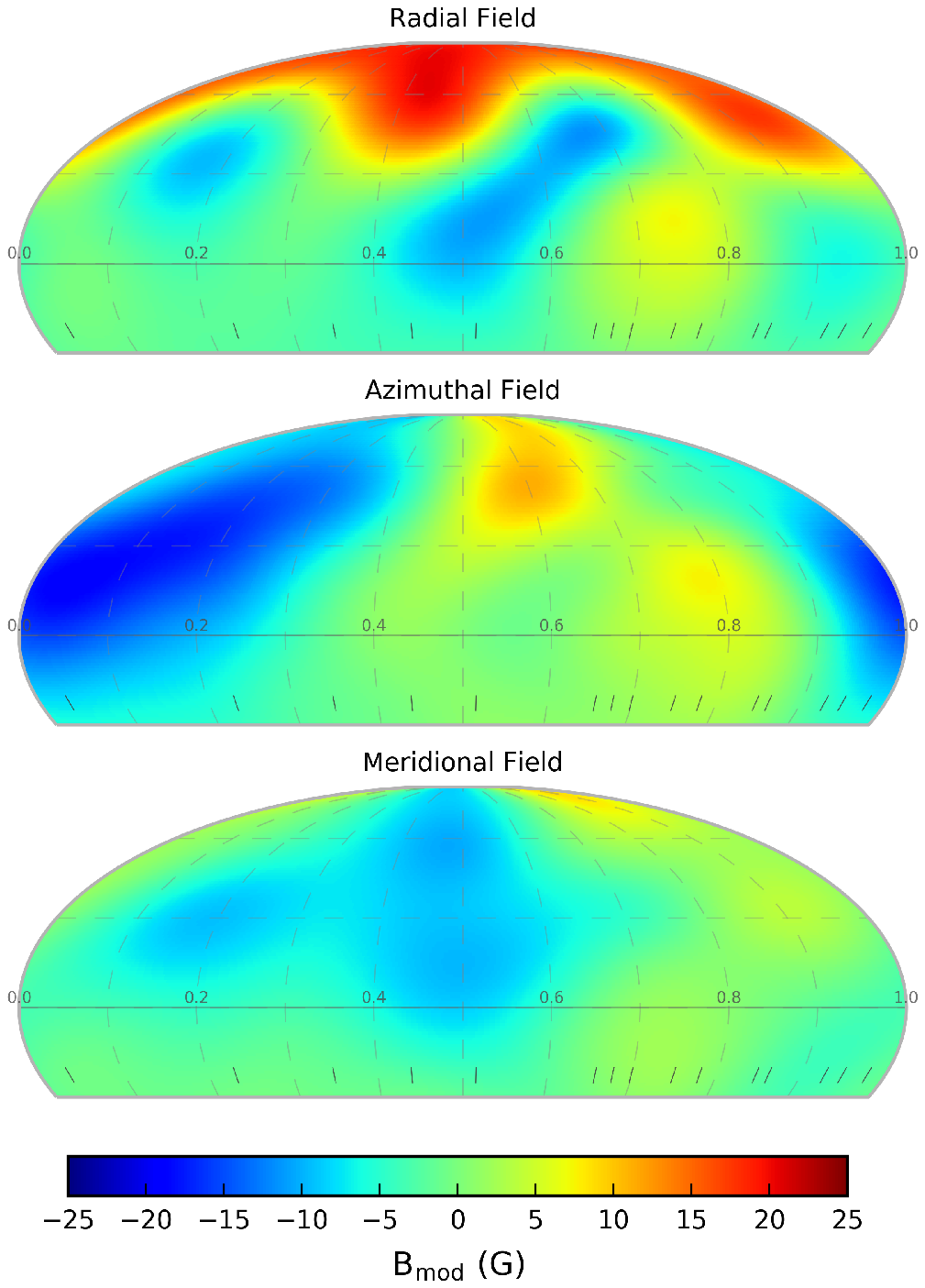}} &
  \imagetop{\includegraphics[height=7.5cm,width=3.3cm,bb = 54 -36 520 780,clip]{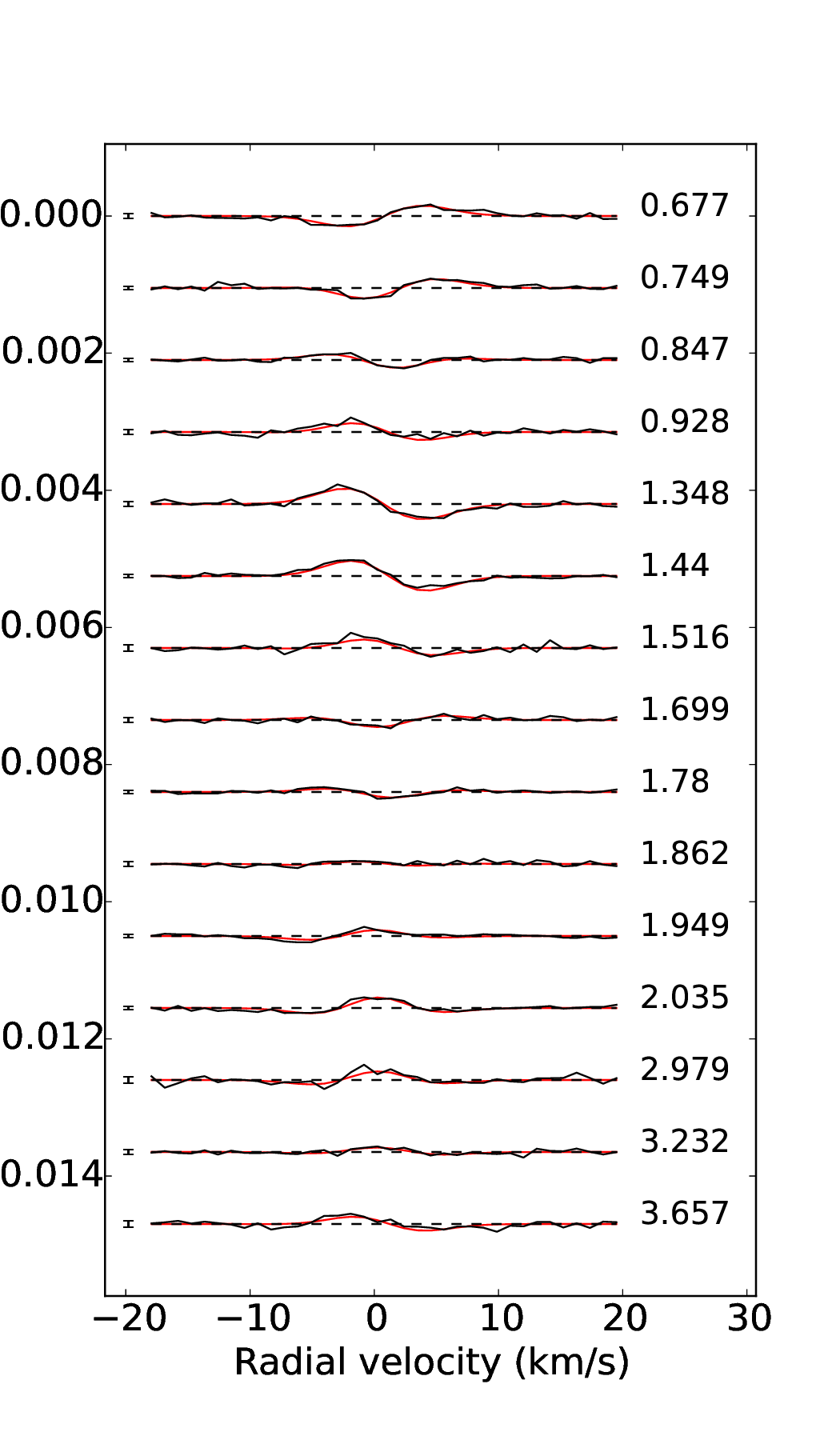}} \\
  \imagetop{\includegraphics[height=7.5cm,width=3.4cm,bb = 54 -36 520 780,clip]{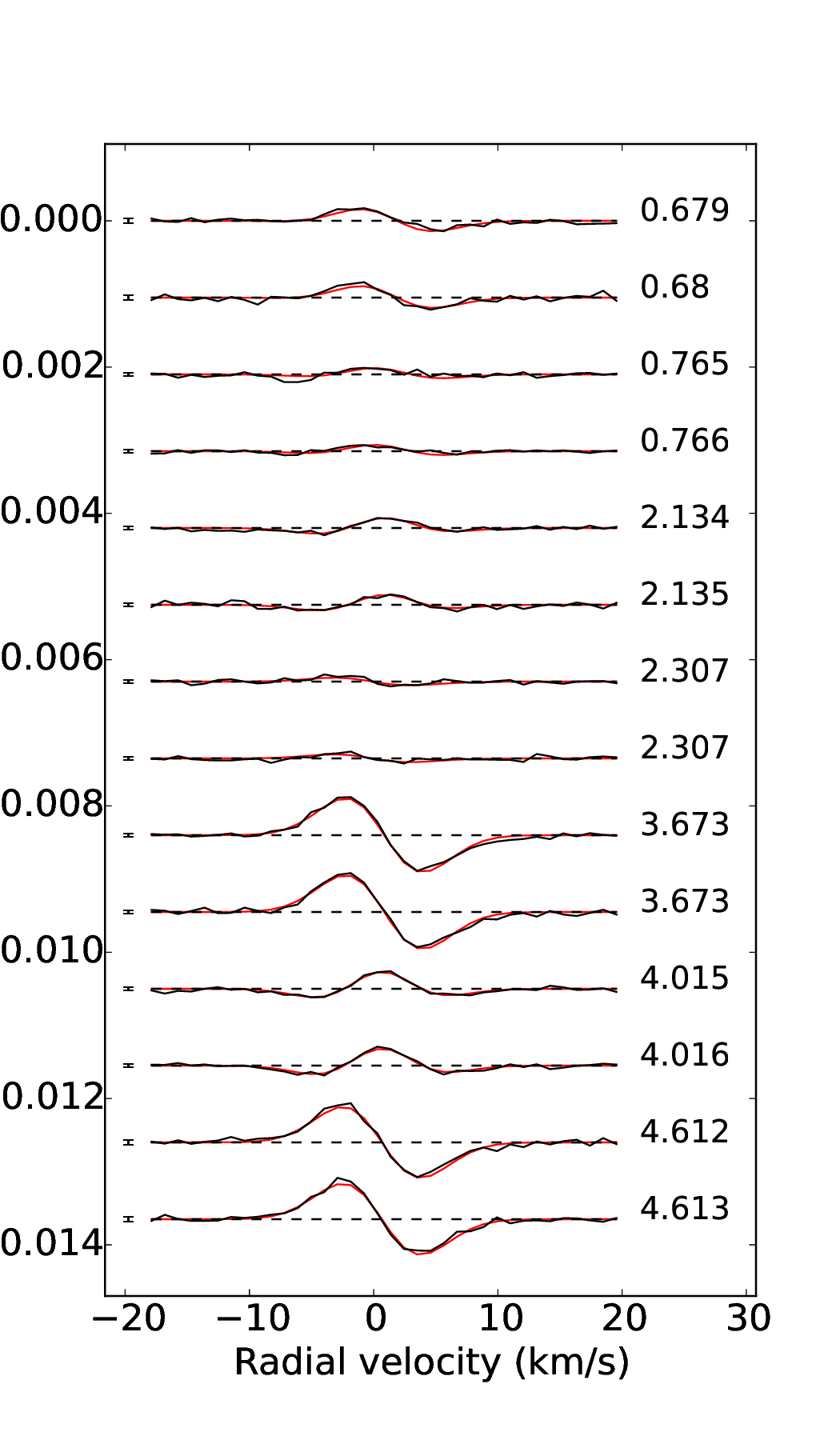}} &
  \imagetop{\includegraphics[height=7.0cm,width=0.29\textwidth]{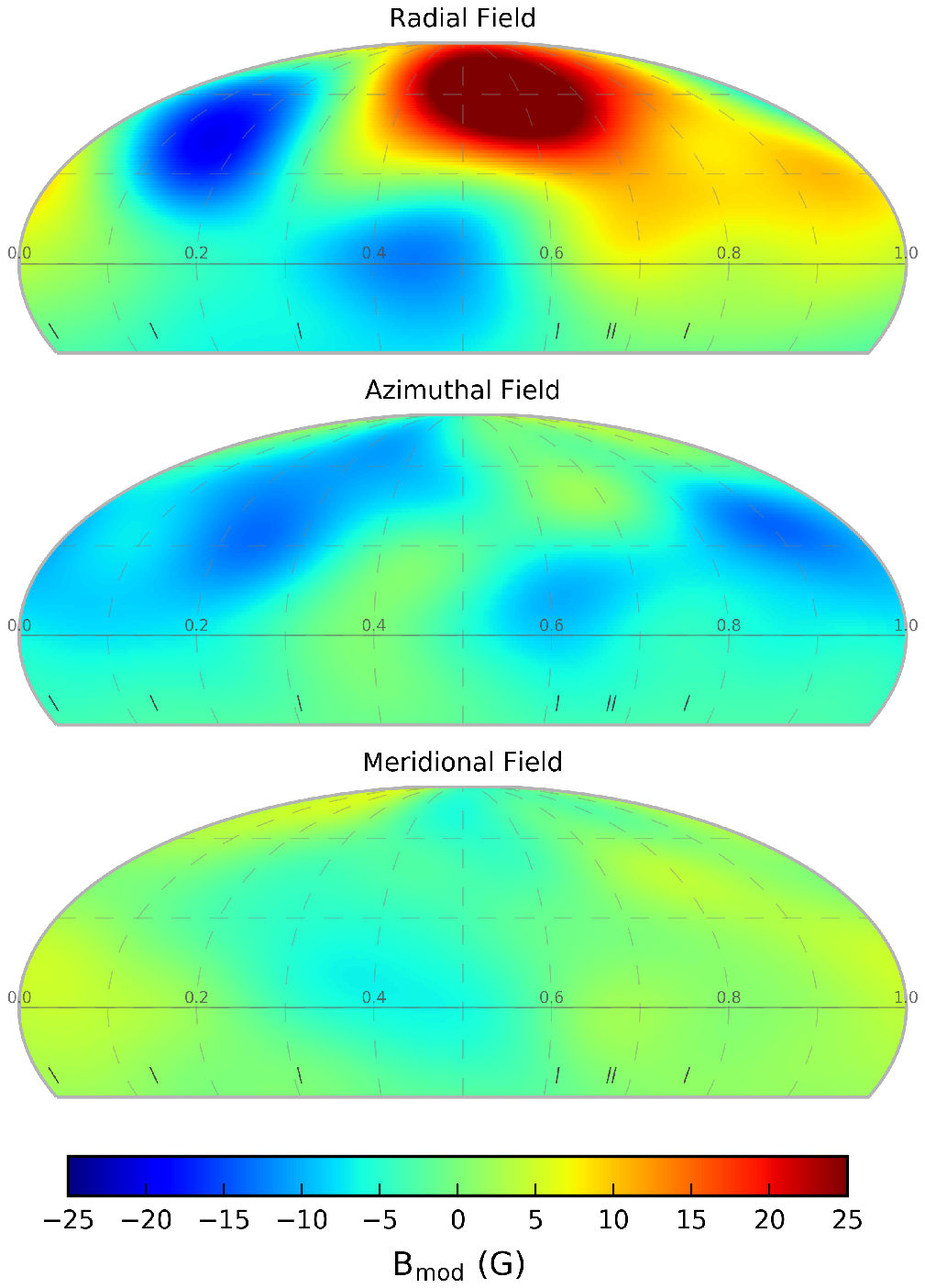}} & 
\end{tabular}
\caption{Magnetic field maps of $\epsilon$ Eridani reconstructed for 2014.71, 2014.84 and 2014.98, shown with the {\em{Stokes V}} fits to the sides (ordered left to right and top to bottom).  For each image the magnetic field projection is shown in terms of radial (upper), azimuthal (middle), and meridional (lower) field components, where red indicates positive polarity and blue negative polarity.  The magnetic field strength is in Gauss where for each map the scale is identical (Bmax = 25 G).  The tick marks at the top of each radial field map indicate the observational phases used to reconstruct the large-scale magnetic field geometry. The {\em{Stokes V}} profiles are plotted separated by a constant value for clarity.}  
\protect\label{f-magmaps} 
\end{figure*}

The reconstructed large scale magnetic field is shown in the central panels of Figure~\ref{f-magmaps}.  The observed and the modelled {\em{Stokes V}} LSD profiles are shown to the sides of the magnetic maps. Over the 5 month timespan of the observations there is a significant evolution of the large-scale magnetic field topology of $\epsilon$ Eridani.  The total observations were divided up into three epochs to avoid the presence of large gaps without observations, resulting from poor weather conditions.  The division of the observations into the maps was tested for different combinations of observations (e.g. five maps versus three maps) and the result was comparable to the maps presented in Figure~\ref{f-magmaps}, though a slightly lower $\chi^2$ was obtained for the data set divided into three maps.   We extensively tested the phase coverage of the maps by assigning random phases to the epochs of observation which resulted in a very similar configuration of magnetic features.  We determine the differential rotation of the magnetic features, as described in \cite{Jeffers2014}, using the first two epochs which was calculated to be $\Omega_{eq}$ = 0.593 rad d$^{-1}$,  $\delta\Omega$ = 0.151 rad d$^{-1}$, which is equivalent to $P_{eq}$ = 10.58d and the $P_{pole}$ = 14.21d.  This is in agreement with our previous measurements of differential rotation for $\epsilon$ Eridani using magnetic features \cite{Jeffers2014}.  Other differential rotation measurements for $\epsilon$ Eridani have been measured using photometric data taken with the MOST satellite where values of 11.35 days, 11.55 days are measured for two different spots \citep{Croll2006}, or $P_{min}$ = 11.04 days and $P_{max}$ = 12.18 days by \cite{Donahue1996}.  While all of these values broadly agree, the differences can be explained by each method measuring different features, e.g. plage regions, photometry and magnetic features which do not necessarily probe the same depth in the stellar atmosphere or stellar latitudes.   Since there are a range of vsini values in the literature we also reconstructed the ZDI maps for an inclination of 30$^\circ$ corresponding to the lowest vsini value of 1.7$\pm$0.3 km s$^{-1}$ \citep{Saar1997}.  To fit the Stokes V profiles required a higher $\chi^2$ value to fit the images (1.12 compared to 0.99 for map1).  In the reconstructed maps, high latitude features are slightly less extended in latitude and longitude, though not more than 5-10 degrees, for the radial component.  The meridional and azimuthal components show a slight increase in the strength of the features of the order of 10\%.  We conclude that the use of a lower inclination does not have any significant impact on the reconstructed features in the ZDI maps.

Even though the phase coverage of the last map is not optimal we note that there is significant evolution of the {\em{Stokes V}} profiles between the same rotation phase (phase 0.67), which can be taken into account when differential rotation is included in the reconstruction of the maps.  Additionally, for map 3, we note that the {\em{Stokes V}} profiles have the largest amplitudes and that a simple magnetic field topology is expected.

\subsubsection{Radial component}

The radial component of $\epsilon$ Eridani's large-scale magnetic field is dominated by a persistent and large polar magnetic spot of positive polarity which varies in strength and shape.  The strength of the region is highest in 2014.98 (map 3), and weakest in 2014.84 (map 2) while the global shape ranges in latitude from 30$^\circ$ (2014.84) to 60$^\circ$ (2014.98).  The strongest evolution of the magnetic field is from epoch 2014.84 (map 2) to epoch 2014.98 (map 3) where the polar spot has increased in strength and evolved into a large spot that is no longer centered at the pole and extends to much lower latitudes.  This evolution of magnetic features is also evident in the corresponding {\em{Stokes V}} profiles.  Additionally, there are weaker magnetic spots with negative polarity that are likely to be evolving from one epoch to the next.  The strength of the negative regions is highest at epochs 2014.98 (map 3) and 2014.71 (map 1) and very weak in 2014.84 (map 2).

\subsubsection{Azimuthal component}

The azimuthal component of $\epsilon$ Eridani's large-scale field appears only very weakly with small hints of magnetic spots at epochs 2014.71 (map 1) and 2014.98 (map 3).  The azimuthal field is strongest at epoch 2014.84 (map 2) where a large magnetic spot of negative polarity extending from phases 0 to 0.3.    

\subsubsection{Meridional component}

The determination of the meridional field is often difficult due to the cross talk  between the meridional and the radial field.  As discussed by \cite{Donati1997a}, the cross talk between the radial and meridional fields primarily effects magnetic features at low latitudes, which implies that the higher latitude features are reliably reconstructed.  This is evident in the magnetic maps, where the reconstructed negative polarity magnetic features at latitudes 0$^\circ$ to 50$^\circ$ in the meridional field maps are mirrored in the radial field maps (for all maps).  The presence of high latitude meridional field in Map 2 at phase 0.5, is considered to be reliable.  An additional important consideration is if the cross-talk is from radial to meridional or vice versa.  As discussed by \cite{Donati1997a}, for stars with low inclinations ($i< 30^\circ$), the cross talk is from radial to meridional while for higher inclinations  ($i> 50^\circ$), the cross talk will be from meridional to radial.  Since the adopted inclination of  $\epsilon$ Eridani  is 46$^\circ$, it lies between these two possibilities and it is not possible to conclude in which direction the cross talk occurs.

\subsection{Magnetic energy}

\begin{figure} 
\begin{center}
\includegraphics[scale=0.3]{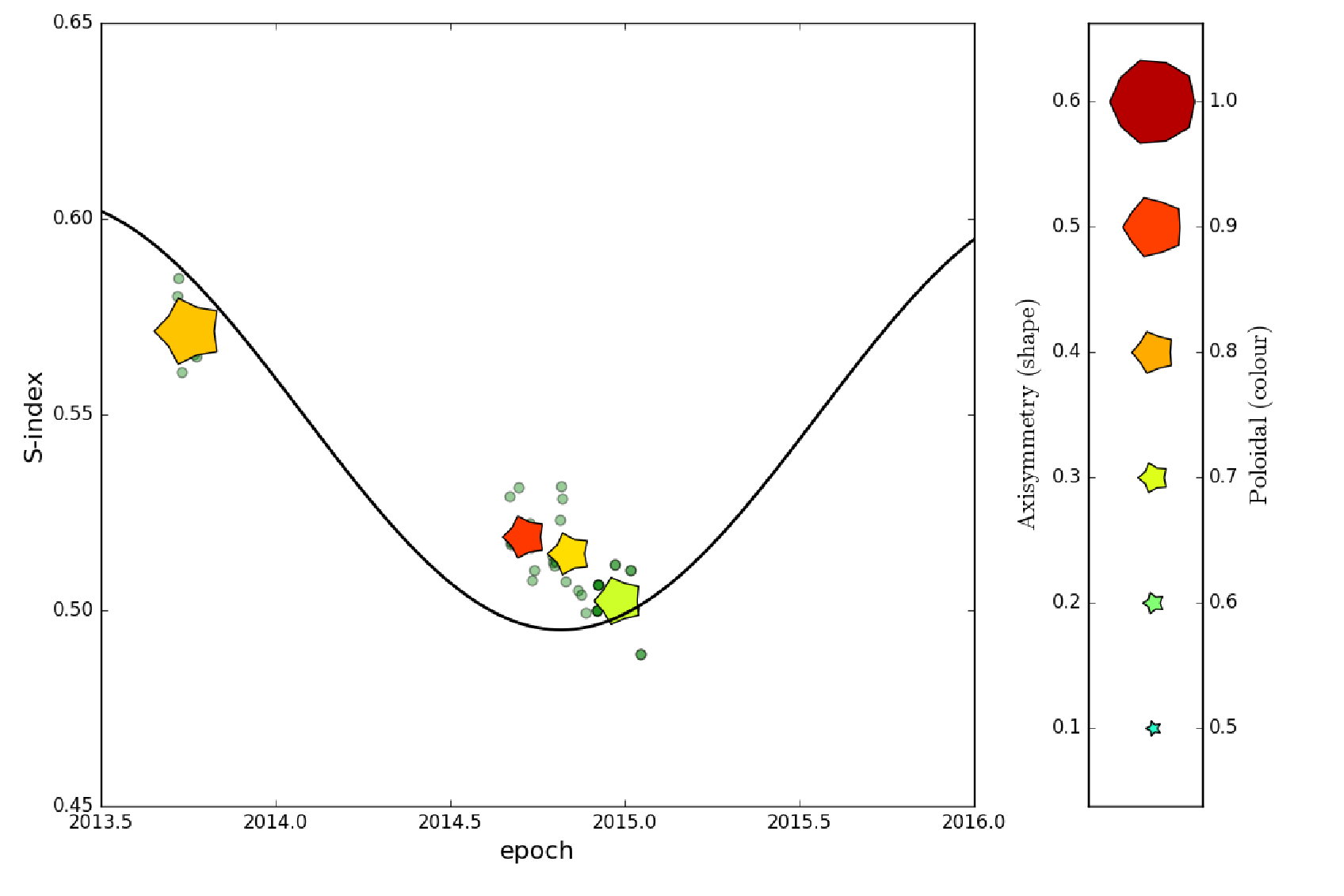}
\caption{The evolution of $\epsilon$ Eridani's large-scale field during S-index minimum.
The symbol shape indicates the axisymmetry of the field (non axisymmetric by pointed star shape and axisymmetric by decagon), the colour of the symbol indicates the proportion of poloidal (red) and toroidal (blue) components of the field and the symbol size indicates the magnetic field strength.  Additionally, S-index points before and after the activity minimum are included (from Jeffers et al. 2014) and unpublished data. The black line indicates the sinusoidal period of 2.95 days and epoch of S-index minimum using the values of Metcalfe et al. (2013).}

\protect\label{f-confusogram} 
\end{center} 
\end{figure}

Over a period of approximately 5 months, $\epsilon$ Eridani's large-scale field evolves (as shown in Fig. 2 and Table 3) with decreasing S-index.  The most dramatic changes are seen in the rapid emergence of an axisymmetric toroidal field.  This is indicated by the colour of the points in Figure~\ref{f-confusogram} changing from red in 2014.71 (map 1) to green in 2014.98 (map 3).  The field is notably more complex than a simple dipole at all epochs with significant amounts of the magnetic energy being contained in higher order modes.  The poloidal component is approximately 50\% dipolar (with values ranging from 43\% in 2014.71 to 56\% in 2014.84), with additional contributions from the quadrupolar (with an average of 20\%) the octupolar component (which are typically of the order of 20\%) and higher order modes $l > 3$.  The axisymmetry of the large-scale field is quite constant with an average value of 35\%.

\section{Discussion}

The large-scale magnetic field geometry of $\epsilon$ Eridani has been shown to be highly variable over its 2.95 year chromospheric activity cycle when observed at yearly epochs \citep{Jeffers2014}.  To investigate the evolution on shorter timescales of the magnetic field geometry of $\epsilon$ Eridani, we secured observations over a period of five successive months, from September 2014 to January 2015, spanning $\epsilon$ Eridani's chromospheric activity minimum.  The large-scale magnetic field is shown to vary on a timescale of months with the first magnetic map reconstructed for epoch 2014.68 (map 1) showing significant evolution compared to the final map reconstructed for epoch 2014.98 (map 3).  The large-scale magnetic field geometry is predominantly poloidal (ranging from 74\% to 84\%) throughout $\epsilon$ Eridani's activity minimum and in contrast to the rest of its activity cycle where it is also seen to have a strong toroidal components \citep{Jeffers2014}.  The poloidal field is not a simple dipole but is quite complex with significant fractions in higher order modes such as quadrupolar and octupolar modes.

The reconstructed complex poloidal field is in contrast to the Sun where its large-scale field is a simple dipole at activity minimum and becomes complex at activity maximum \citep{DeRosa2012}. On the Sun, the dipolar and quadrupolar modes vary in anti-phase, such that after cycle minimum, the quadrupolar mode grows as the number of spots increases, reaching a maximum at cycle maximum \citep{DeRosa2012}.  However, on $\epsilon$ Eridani, we observe a slight decrease in the quadrupolar mode at this phase, accompanied by a slight growth in the dipole mode.  The changes are small, and a longer term study is needed to confirm if this is indeed the pattern that characterises the cycle.  The most significant change in $\epsilon$ Eridani's large-scale magnetic field is in the toroidal component of the axisymetric field which evolves from 5\% to 72\% precisely at the emergence of $\epsilon$ Eridani from its activity minimum and shows a strong axisymmetry, just like the Sun's poloidal field.   Currently, there is limited information on the long-term evolution of the Sun's toroidal field as only a few years of vector data are available ~\citep{Gosain2013, Vidotto2016}. Within this small time window, however, the solar  toroidal field was much weaker than on $\epsilon$ Eridani.   A longer term comparison of the variation of the toroidal field over the stellar magnetic cycle may shed some light on the nature of the magnetic cycle.

\begin{table*}
\caption{The fraction of the large-scale magnetic energy reconstructed in the toroidal and poloidal field components; the fraction of the poloidal field in the dipolar ($\ell$=1), quadrupolar ($\ell$=2) and octupolar ($\ell$=3) components; and the fraction of the energy stored in the axisymmetric component ($m$=0) (e.g. the fraction of the poloidal field that is axisymmetric and the fraction of the toroidal field that is axisymmetric. )}
\vspace{-0.4cm}
\protect\label{t-mag_en}
\begin{center}
\begin{tabular}{c c c c c c c c c c c c c }
\hline
\hline
Epoch & Bmean & Bmax & Toroidal & Poloidal & Dipolar & Quadrupolar & Octupolar & $\ell>$3 & Axisymmetric & Poloidal & Toroidal \\ 
 & (G) & (G) & (\% tot) & (\% tot) & (\% pol) & (\% pol) & (\% pol) & (\% pol) & (\% tot) & (\% axi) & (\% axi) \\
\hline
2014.71 & 8$\pm$1  & 28  & 16  & 84  & 43  & 25  & 19 & 13 & 35  & 41  & 5 \\ 
2014.84 & 8$\pm$1  & 20  & 23  & 77  & 56  & 14  & 14 & 16 & 32  & 28  & 43 \\ 
2014.98 & 10$\pm$1  & 33  & 26  & 74  & 50  & 19  & 22 & 9 & 40  & 29  & 72 \\ 
\hline
\hline
\vspace{-1.0cm}
\end{tabular}
\end{center}
\end{table*}

Another star that has been monitored as part of the BCool survey and that has stellar parameters very similar to $\epsilon$ Eridani is 61 Cyg A \citep{BoroSaikia2016}, which also exhibits a solar-like magnetic cycle.  At activity minimum, the large-scale field of 61 Cyg A is also a simple dipole like the solar case, showing that it is not a limitation of ZDI that we do not see similar behaviour for $\epsilon$ Eridani.  The poloidal field of $\epsilon$ Eridani is more complex at activity minimum compared to 61 Cyg A and to the Sun.  The S-index cycle of 61 Cyg A is 7.2$\pm$1.3 years long \citep{BoroSaikia2016} and its stellar parameters are more similar to $\epsilon$ Eridani than the Sun's.  The mass of 61 Cyg A is 0.66 M$_\odot$ \citep{Kervella2008} which is slightly smaller than $\epsilon$ Eridani's mass of 0.7 M$_\odot$ and given their low vsini values, the ZDI technique has a similar resolving power for both stars.  The evolutionary state of the two stars is similar with $\epsilon$ Eridani having an age that is approximately 7\% of its main-sequence lifetime compared to 14\% for 61 CygA \citep[calculated using the stellar evolution models of][]{Pols1998}.  The main difference is the rotation periods of the two stars, with 61 Cyg A having a rotation period that is approximately three times as long, 35.4 days compared to $\epsilon$ Eridani's 11.68 days.

The strength of the mean magnetic field remains constant despite the changing field geometry.  In contrast to this, the maximum strength varies from 20 G to 33 G.  Since this is the strength in the large-scale component, there is likely to be additional contributions from small-scale component that remains undetected with techniques such as ZDI.  Evidence for additional small scale field is shown by magnetic field measurements using {\em{Stokes I}} (unpolarised) line broadening which measures the strength of both the large-scale and the small-scale field.  These values are typically of the order of 127 G \citep{Valenti1995} or 165 G \citep{Rueedi1997}.  As found by previous authors, the large-scale field is of the order of 10\% of the total field.  The maximum and mean magnetic field strengths measured in 2014.9 (map3) are comparable to previous values reconstructed at activity minimum \citep[][epoch 2011.81]{Jeffers2014}.  Additionally at epoch 2011.81, the large-scale field geometry has a large magnetic spot with positive polarity which is very similar to the map reconstructed in 2014.9 (map 3).  The two maps have the same fraction of poloidal field (74\% for both maps), which has a much more complex structure in 2014.9 compared to 2011.81.  The main difference between the two maps is the fraction of axisymmetric field which for epoch 2014.9 comprises 40\% and 2011.81 comprises 63\%.  However, over the three epochs of this analysis there are dramatic changes in the geometry of the axisymmetric field which evolves from 41\% to 29\% for the poloidal component and 5\% to 72\% for the toroidal component.  

\section{Conclusions}
The high cadence observations of $\epsilon$ Eridani's large-scale magnetic field geometry shows that the large-scale magnetic field geometry evolves on a timescale of months with a dramatic increase in the toroidal component of the axisymmetric field at the emergence from its activity minimum.  The large-scale field also shows a predominantly poloidal component that is surprisingly complex when compared to the Sun at activity minimum.  Our results show that the magnetic field of solar-type stars can be quite different from the Sun even when they exhibit clear chromospheric activity cycles. 

\section*{Acknowledgments}

We would like to thank M.Mengel, B. Carter, C. Folsom and V. See for valuable contributions in discussing the results of this paper, and G. Anglada Escude and A. Reiners for their contributions to the observing proposal.  JRB was supported by the UK Science and Technology Facilities Council (STFC) under the grant ST/L000776/1. The research leading to these results has received funding from the European Community's Seventh Framework Programme (FP7/2013-2016) under grant agreement number 312430 (OPTICON). 

\bibliographystyle{mnras}
\bibliography{iau_journals,epseri}

\bsp	
\label{lastpage}
\end{document}